\documentclass[usenatbib]{mn2e}
\usepackage{times}
\usepackage{supertabular,multicol}

\newcommand{\rms}{$20$~seconds}

\usepackage{color}
\usepackage{graphicx}
\begin{document}

\title[TraMoS Project III: Analysis of 38 transit observations of
  WASP-4b exoplanet.]{TraMoS project III: Improved physical
  parameters, timing analysis, and star-spot modelling of the
  WASP-4b exoplanet system from 38 transit observations.}

\author[S. Hoyer et al.] {S.~Hoyer$^{1,5,6}$,
  M.~L\'opez-Morales$^{2}$, P.~Rojo$^1$, V. Nascimbeni$^{3,4}$,
  S.~Hidalgo$^{5,6}$, \newauthor N.~Astudillo-Defru$^{1,7}$,
  F.~Concha$^1$, Y.~Contreras$^1$, E.~Servajean$^1$,
  T.C. Hinse$^{8,9}$.\\ $^1$ Astronomy Department, Universidad de
  Chile, Casilla 36-D, Santiago de Chile, Chile.\\ $^2$
  Harvard-Smithsonian Center for Astrophysics, 60 Garden Street,
  Cambridge, MA 02138, USA.\\ $^3$ Dipartimento di Fisica e
  Astronomia, Universit\`a degli Studi di Padova, Vicolo
  dell'Osservatorio 3, 35122 Padova, Italy. \\ $^4$ INAF -
  Osservatorio Astronomico di Padova, Vicolo dell'Osservatorio 5,
  35122 Padova, Italy. \\ $^5$ Instituto de Astrof\'isica de Canarias,
  Via L\'actea s/n, E38200 La Laguna, Tenerife, Canary Islands,
  Spain. \\ $^6$ Department of Astrophysics, University of La Laguna,
  Via L\'actea s/n, E38200 La Laguna, Tenerife, Canary Islands,
  Spain. \\ $^{7}$UJF-Grenoble 1/CNRS-INSU, Institut de Planétologie
  et d'Astrophysique de Grenoble (IPAG) UMR 5274, Grenoble, F-38041,
  France \\ $^8$ Korea Astronomy and Space Science Institute, 776
  Daedukdae-ro, Yuseong-gu, 305-348 Daejeon, Republic of
  Korea. \\ $^9$ Armagh Observatory, College Hill, BT61 9DG, United
  Kingdom.}

\maketitle

\begin{abstract}

We report twelve new transit observations of the exoplanet WASP-4b
from the Transit Monitoring in the South Project (\textit{TraMoS})
project. These transits are combined with all previously published
transit data for this planet to provide an improved radius measurement
of $R_p= 1.395 \pm 0.022~R_{jup}$ and improved transit ephemerides. In
a new homogeneous analysis in search for Transit Timing Variations
(TTVs) we find no evidence of those with {\it RMS} amplitudes larger
than \rms ~over a 4-year time span.  This lack of TTVs rules out
the presence of additional planets in the system with masses larger
than about $2.5 ~M_{\earth}$, $2.0~M_{\earth}$ and $1.0~M_{\earth}$
around the 1:2, 5:3 and 2:1 orbital resonances.  Our search for the
variation of other parameters, such as orbital inclination and transit
depth also yields negative results over the total time span of the
transit observations.  Finally we perform a simple study of stellar
spots configurations of the system and conclude that the star
rotational period is about 34 days.

\end{abstract}


\section{Introduction}

Since the discovery of the first extrasolar planet around the Sun-like
star 51-Peg via radial velocities \citep{mayorandqueloz95},
%and the discovery of the first transiting planet around the nearby solar-type
%star HD~209458 \citep{Charbonneau2000,Henry2000} 
a number of systematic extrasolar planet searches have spread
adopting a wide variety of techniques, of which the radial velocity
method is still the most prolific approach.  The transit technique is
currently the second most successful, with the detection of over 290
systems with confirmed planet detection\footnote{The extrasolar planet
  encyclopaedia: http://www.exoplanet.eu}.  Of these, most are
Hot-Jupiters (Jupiter-mass objects with orbital periods of a few
days).  Just recently, space missions such as Kepler and Corot have
started to expand transit discoveries to planets smaller than
$50~M_{\earth}$.

Transiting planets provide a wealth of information about their
systems.  For instance, transits are currently the only tool to measure the
planet-to-star radius ratio and orbital inclination.  Combined with
radial velocity results, those parameters allow the determination of
the absolute mass of the planet and its mean density.  Another type of
study that can be conducted via transiting planets is the search of
{\it unseen} companions in the system.  Those companions introduce
variations in the orbital period of the transiting planet
\citep{miralda02,Agol05,holman05}, which can be detected by monitoring
the systems in search for Transit Timing Variations (TTVs).  This TTV
technique has the potential of finding planets in the Earth-mass
regime and even exomoons \citep{Kipping09a}.

%The transit technique is currently the only way to measure the
%planet-to-star radius ratio and orbital inclination.  Combined with
%radial velocity observations it allows the determination of the mass
%and consequently the mean density of the planet.

%Among the wide variety of studies that can be conducted with
%transiting exoplanets, it has been proved that the presence of
%\textit{unseen} orbital companions can produce changes in the orbital
%period of transiting exoplanets \citep{miralda02,agol05,holman05}.
%Detection of additional planets in the system can be done by
%monitoring for this Transit Timing Variations (TTVs).  The TTV
%technique has the potential of finding planets in the Earth-like mass
%regime or exomoons \citep{Kipping09a}.   

In addition, \cite{Silva-Valio2008} and \cite{Sanchis-Ojeda2011} have
pointed out that observations of star-spot occultations during
closely-spaced transits can be used to not only estimate the rotation
period of the host star, but also to measure alignment differences
between the rotation axis of the star and the orbital axis of the
planet.

%Recently, the Kepler Space Mission \citep{Borucki10} has given the
%most remarkable TTVs discoveries: Kepler-9, a system with two
%Saturn-like transiting planets with TTVs of amplitudes of tens of
%minutes \citep{Holman10}; Kepler-11, a system with six transiting
%planets (all of them show TTVs) \citep{Lissauer11}; Kepler-18, a
%system with three planets confirmed by TTVs \citep{Cochran11}; and
%Kepler-19, a transiting planet where the presence of an additional
%orbiting body was inferred from its TTVs \citep{Ballard11}.  All these
%exoplanets have masses in the range of 2 to 80 $M_{\earth}$.  Based on
%an statistical study of the first list of exoplanet candidates from
%Kepler, between $11\%-20\%$ of the suitable targets for timing
%analysis show some evidence of TTVs \citep{Ford.Kepler.TTV.2011}.

As part of the Transit Monitoring in the South (\textit{TraMoS})
project \citep{Hoyer-ogletr111-2011, Hoyer-wasp5-2012}, we are
conducting a photometric monitoring survey of transits observable from
the Southern Hemisphere.  The aim of this project is to perform a
careful and homogeneous monitoring of exoplanet transits trying to
minimize systematics and reduce uncertainties in the transit
parameters, such as the transit mid-time, following the approach of
using high-cadence observations and the same instruments and setups.

In the framework of the \textit{TraMoS} project we present twelve new
transit observations of the exoplanet WASP-4b.
%collected between August 2008 and September 2011.  
This was the first exoplanet detected by the WASP-South survey in
2008.  A that time, \citet[hereafter W08]{Wilson08} reported a
Hot-Jupiter ($P=1.34$~days) with a mass of
$M_{P}=1.22^{+0.09}_{-0.08}~M_J$ and a planetary radius of
$R_{p}=1.42^{+0.07}_{-0.04} ~R_{J}$ orbiting a G7V southern star.
This discovery paper included WASP photometry, two additional transit
epochs (observed in September 2007), and radial velocities
measurements.

\citet[hereafter G09]{Gillon.WASP4WASP5.2009}, added to the follow-up
of this exoplanet a \textit{VLT/FORS2} light curve observed in October
2007 with a \textit{z-GUNN} filter.  Using a reanalysis of the W08
data they found no evidence of period variability.  \citet[hereafter
  W09]{Winn.WASP4.2009}, presented two new high-quality transits
observed in 2008 with the Baade Telescope (one of the twin 6.5-m
Magellan telescopes at Las Campanas Observatory) using a
\textit{z-band} filter.  Four new transit epochs were reported shortly
after by \citet[hereafter S09]{Southworth.WASP4.2009}, with the 1.54-m
Danish Telescope at La Silla Observatory using a \textit{Cousins-R}
filter during 2008.  \citet[hereafter S011]{Sanchis-Ojeda2011}, using
four new transit light curves observed during 2009, interpreted two
anomalies in the photometry as starspot occultations by the planet and
concluded from that result that the stellar rotation axis is nearly
aligned with the planet's orbital axis.  This result agrees with the
observations of the Rossiter-McLaughlin effect for this system by
\cite{Triaud2010}.  Later, two new transits of WASP-4b were reported
by \citet[hereafter D11]{Dragomir11}, with data from the 1-m telescope
at Cerro Tololo Inter-American Observatory (\textit{V-band} and
\textit{R-band} filter).  Most recently, Nikolov et al. (2012,
hereafter N12) observed three transits simultaneously in the {\it
  Sloan g', r', i' and z'} filters with the Gamma Ray burst Optical
and Near-infrared Detector (GROND) at the MPG/ESO-2.2 m telescope at
La Silla Observatory.

%Most recently, \citet[hereafter N12]{Nikolov-wasp4-2012}
%with data of the Gamma Ray burst Optical and Near-infrared Detector
%(GROND) at the MPG/ESO-2.2 m telescope at La Silla Observatory
%observed three transits simultaneously in the {\it Sloan g',r',i'} and
%{\it z'} filters.

%\footnote{By the time of the writing of this
%  work three new transits were reported by \cite{Nikolov-wasp4-2012}
%  with data of the Gamma Ray burst Optical and Near-infrared Detector
%  (GROND) at the MPG/ESO-2.2 m telescope at La Silla Observatory
%  (observed simultaneously in the {\it Sloan g',r',i'} and {\it z'}
%  filters).  We did not model these light curves but we used their
%  reported midtimes in part of the timing analysis (See section
%\ref{ttv}).}.

In this work we present twelve new transits observations.  We combined
these new light curves with all the previously available light curves
(twenty-six additional light curves) and reanalyzed them to provide a
new homogeneous timing analysis of the transits of WASP-4b and to
place stronger constraints to the mass of potential perturbers in the
main orbital resonances with this planet. We also search the entire
dataset for signs of stellar spots that would help improve the
conclusions of S011.

In Section \ref{observaciones} and \ref{reduccion} we describe the new
observations and the data reduction. Section \ref{fiteo} details the
modelling of the light curves and in Section \ref{ttv} we present the
timing analysis and discuss the mass limits for \textit{unseen}
perturbers.  In section \ref{spots} we discuss the occultations of
stellar spots by the planet.  Finally, we present our conclusions in
Section \ref{conclusiones}.

\section{Observations} \label{observaciones}

%% FIGURE 1
\begin{figure*}
\vspace{10cm} 
\includegraphics{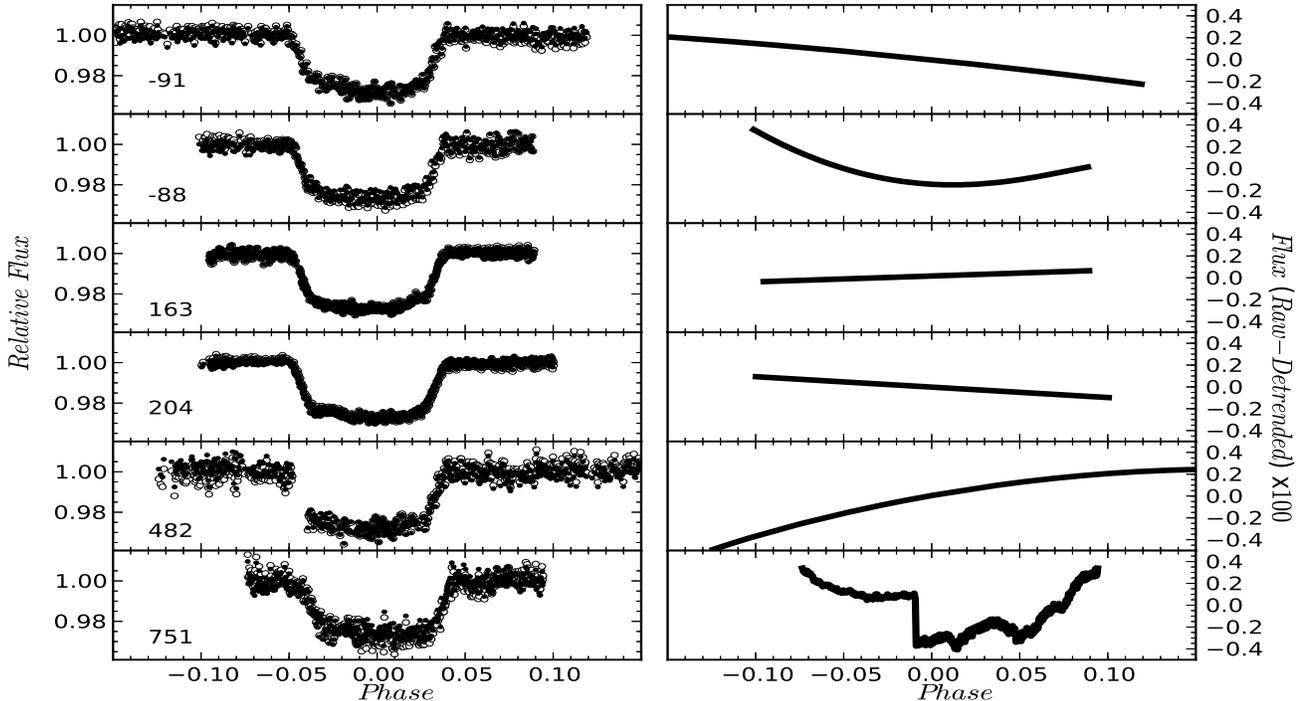}
\caption{Left: Detrended light curves (black points) plotted
  over the raw light curves (gray points) to illustrate the amount of
  detrending applied before the transit modelling using MCMC. The
  epoch of each transit is shown in the bottom left of each
  panel. Right: The flux difference (x 100 times) between the
  raw and detrended light curves. }
\label{rawvsdetrended}
\end{figure*}

\subsection{The instruments}

As mentioned before, the~\textit{TraMoS} project has undertaken a
photometric campaign to follow-up transiting planets observable from
the Southern Hemisphere.  Our goal is to use high cadence observations
minimizing change of instruments to reduce systematics and therefore,
based on an homogeneous analysis, obtain the most precise values of
the light curve parameters, such as the central time of the transit,
the orbital inclination and the planet radius, among others.

The observations we present in this work were performed with the
Y4KCam on the SMARTS 1-m Telecope at Cerro Tololo Inter-American
Observatory (CTIO) and with the SOAR Optical Imager (SOI) at the 4.2-m
Southern Astrophysical Research (SOAR) telescope in Cerro Pach{\'on}.
The epochs of four of the transits we present here coincide with
previous published data (see Table \ref{tabla-obs}).

We have taken advantage of the $20\times20$ squared arcminute of
Field of View (FoV) of the Y4KCam, which is a $4064\times4064$ CCD
camera with a pixel scale of 0.289 arcsec pixel$^{-1}$, which despite its
large dimensions allows to use a readout time of only
$\sim16/5$~sec when using the 2x2/4x4 binning mode (compared with the
$46$~sec of the unbinned readout time).  The SOI detector is composed
of two E2V mosaics of $4096\times2048$ pixels with a scale of 0.077
arcsec pixel$^{-1}$. The SOI has a FoV of $5.2\times5.2$ squared
arcminutes and allows a readout time of only $\sim11$~sec after
binning 2x2 ($20.6$~sec is its standard readout time).

As part of the \textit{TraMoS} project we have observed a total of 12
transits of WASP-4b, between August 2008 and September
2011\footnote{In the remaining of the text we refer to each individual
  transit by the epoch number transit, using the ephemeris equation of
  D11.  Transit epoch numbers are listed in the second column of Table
  \ref{tabla-obs}.}.  Three transits were observed with the SOI at the
SOAR telescope and the remaining nine were observed with the Y4KCAM at
the 1-m CTIO telescope. The first ten transits were observed using a
\textit{Bessell I} or \textit{Cousins I} filter. For the 2011 transit
observations we use the 4x4 binning mode of the Y4KCam and a
\textit{Cousins R} filter. The observing log is summarized in
Table~\ref{tabla-obs}.
%($\lambda_{\rm eff}=8665~\AA$ and FWHM=$3914~\AA$). 

All the transits were fully covered by our observations except for
some portions of the $E=754$, $482$ and $-62$ transits that were lost due to
technical failures (see Figure \ref{lc.miscurvas}).  The
before-transit and ingress portions of the $E=-71$ transit were not
observed, but after $phase=-0.034$ (where $phase=0$ is defined as the
phase of the mid-transit) the observation of the transit was done
without interruption.

\begin{table*}
%\tiny
\center
\caption{Details about each of the new transit epoch observation presented in this work.}

\label{tabla-obs}
\begin{tabular}{lccccccc}
\hline
UT date & Epoch$^{b}$  & Telescope/Instrument   & Filter & Binning & Average & Detrending &$K_{\textit{RMS}}$ \\
 & & & & & exptime (sec) &  &  \\ %airmass/linear/quadratic &  \\

\hline
2008-08-23$^{a}$ & -91& 1-mt SMART/Y4KCam & Cousins I & 2x2 & 20  & 3 & 0.78\\
2008-08-23$^{a}$ & -91& SOAR/SOI          & Bessell I & 2x2 & 7  &  -- & --\\ 
2008-08-27 & -88& 1-mt SMART/Y4KCam & Cousins I & 2x2 & 20   & 1+3 & 0.90\\
2008-09-19 & -71& 1-mt SMART/Y4KCam & Cousins I & 2x2 & 20   & -- & -- \\ 
2008-09-23$^{a}$ & -68& 1-mt SMART/Y4KCam & Cousins I & 2x2 & 20  & -- & --\\ 
2008-10-01$^{a}$ & -62& 1-mt SMART/Y4KCam & Cousins I & 2x2 & 26  & --& --\\ 
2009-07-29 & 163& SOAR/SOI          & Bessell I & 2x2 & 8  & 2 & 0.91 \\
2009-09-22 & 204& SOAR/SOI          & Bessell I & 2x2 & 10   & 2 & 0.87 \\
2009-10-28 & 231& 1-mt SMART/Y4KCam & Bessell I & 2x2 & 14  & --  & --\\ 
2010-09-29 & 482& 1-mt SMART/Y4KCam & Cousins I & 2x2 & 14  & 1+3  & 0.96 \\ 
2011-09-24 & 751& 1-mt SMART/Y4KCam & Cousins R & 4x4 & 20  & 1+3  & 0.95 \\ 
2011-09-28 & 754& 1-mt SMART/Y4KCam & Cousins R & 4x4 & 20  & -- & -- \\ 

\hline
\multicolumn{8}{l}{$^{a}$ These epochs have been already observed by other authors. $^{b}$ Epochs are caculated using the ephemeris equation }\\ 
\multicolumn{8}{l}{from \cite{Dragomir11}. $^{c}$ Detrending using (1) airmass, (2) linear or (3) quadratic regression.}\\
\end{tabular}
%\medskip
\end{table*}

\section{Data Reduction} \label{reduccion}

The trimming, bias and flatfield correction of all the collected data
was performed using custom-made pipelines specifically developed for
each instrument.

The \textit{Modified Julian Day} (JD-2400000.5) value was recorded at
the start of each exposure in the image headers.  In the SMARTS
telescope, the time stamp recorded in the header of each frame is
generated by a \textit{IRIG-B GPS time synchronization protocol}
connected to the computers that control the instrument.  The SOAR
telescope data use the time values provided by a time service
connected to the instrument.  We confirmed that these times coincide
with the time of the Universal Time (UT) reference clock within 1
second.
%  values have $\leq 1$~second precision compared with Universal Time clocks.  
The time stamp assigned to each frame corresponds to the
\textit{Julian Day} at the start of the exposure plus 1/2 of the
integration time of each image.

For the photometry and light curve generation we used the same
procedure described in \cite{Hoyer-wasp5-2012}.  That is, we performed
a standard aperture photometry where the optimal sky-aperture
combination was chosen based on the \textit{RMS} minimization of the
differential light curves of the target and the reference stars.  For
this analysis we excluded the ingress and egress portions of the light
curves, i.e. we used only the out-of transit and in-transit data. The
final light curves were generated computing the ratio between the
target flux and the average flux of the best 1 to 3 comparison stars.

In order to detrend the light curves, we searched for correlations of
the out-of-transit flux ($F_{OOT}$) with the X-Y CCD coordinates of
the target, airmass and/or time.  Trends are identified as such when
the correlation coefficient values are larger than the \textit{RMS} of
the out-of-transit points.  To remove the trends, we
  applied linear or quadratic regressions fits, which were subtracted
  from the light curve only when we detect that {\it RMS} of the
  $F_{OOT}$~is improved. Otherwise no subtraction is applied. The {\it
    RMS} after detrending are $3-22\%$ lower than on the raw light
  curves. A posteriori we have checked that doing this detrending
  previous to modelling our light curves does not introduce any bias
  or differences in the estimated parameters and its uncertainties.
  In the last two columns of Table \ref{tabla-obs} we show which
  detrending was applied and the value $K_{{\it RMS}}$, which
  corresponds to the ratio of the $F_{OOT}$~{\it RMS} after and before
  removing the trends to illustrate the improvement in the {\it RMS}
  by the detrending.  Therefore, the values of $K_{RMS}$ indicate how
  much the dispersion of the light curves changes after
  detrending. When no numerical values of $K_{{\it RMS}}$ is shown
  means that no significant systematic trend with any of the
  parameters searched was found. Also, to illustrate the amount of
  detrending applied for each light curve in left panels of Figure
  \ref{rawvsdetrended} we show the detrended over the raw light curves
  and in the right-side of the same figure we also show the
  differences in flux (magnified by 100 times) between the light
  curves before and after detrending.

%  To remove the trends we found, we
%applied linear or quadratic regressions fits, which were subtracted
%from the light curve. \textcolor{red}{The detrending was applied if a
%  significant improvement in the RMS of the $F_{OOT}$~was found.  
%Doing this we reach dispersions in the light
%curves in the range of $0.2 \%$ - $ 0.45\% $ \ojo.

\section{Light Curve Modelling} \label{fiteo}

In a first step we performed a modelling of the 12
  new light curves we present in this work and the 26 available light
  curves of the WASP-4 system from other authors.  The goal of this
  first analysis is to determine transit parameters single values from
  each light curve.  With these values we search for variations in 4 years
  time span in parameters such as the central time of the transits,
  the orbital inclination and/or the transit's depth.  Variations in
  these parameters can be indicative of the presence of a perturbing
  body in the system.  In a second step, we take advantage of the
  information contained in all the light curves by fitting
  simultaneously the transit parameters.  With this global analysis we
  determined the properties of the system.

\subsection{Individual Analysis: searching for parameter variations}\label{individual}

We used the TAP package \citep{TAP} to fit all the available light
curves of WASP-4.  This package allows to fit the analytical models
of \cite{MandelAgol02} based on the Markov Chain Monte Carlo (MCMC)
method which have proved to give the most reliable results compared
with other approaches, particularly in the case of uncertainty estimations of
the fitted parameters \citep[see][section 3.1 and
  3.2]{Hoyer-wasp5-2012}. This point is critical especially for the
timing analysis of the transits.  TAP also implements a wavelet-based
method \citep{Carter09} to account for the red-noise in the light
curve fitting.  This method helps impose more conservative uncertainty
estimates compared to other red noise calculations such as the `\textit{time
averaging}' or `\textit{residual permutation}' methods, in cases where the noise
has a power spectral density (PSD) that varies as $1/f^{\gamma}$ (where $f$ is the frequency and $\gamma$ is a spectral index $> 0$). For all other
types of noise the \cite{Carter09} method gives uncertainty estimations
as good as the other methods \citep[see e.g.][]{Hoyer-ogletr111-2011,Hoyer-wasp5-2012}.

%This method also helps impose more conservative
%uncertainty estimates compared with other red-noise calculators such
%as the 'time averaging' or 'residual permutation' methods (see
%\citealt{Carter09}).

%% FIGURE 2
\begin{figure*}
\vspace{19.5cm} 
%\special{psfile=wasp4.lc.mis12curvas.v3.eps hscale=45   vscale=43 angle=0 hoffset=-15 voffset=-48}
\includegraphics{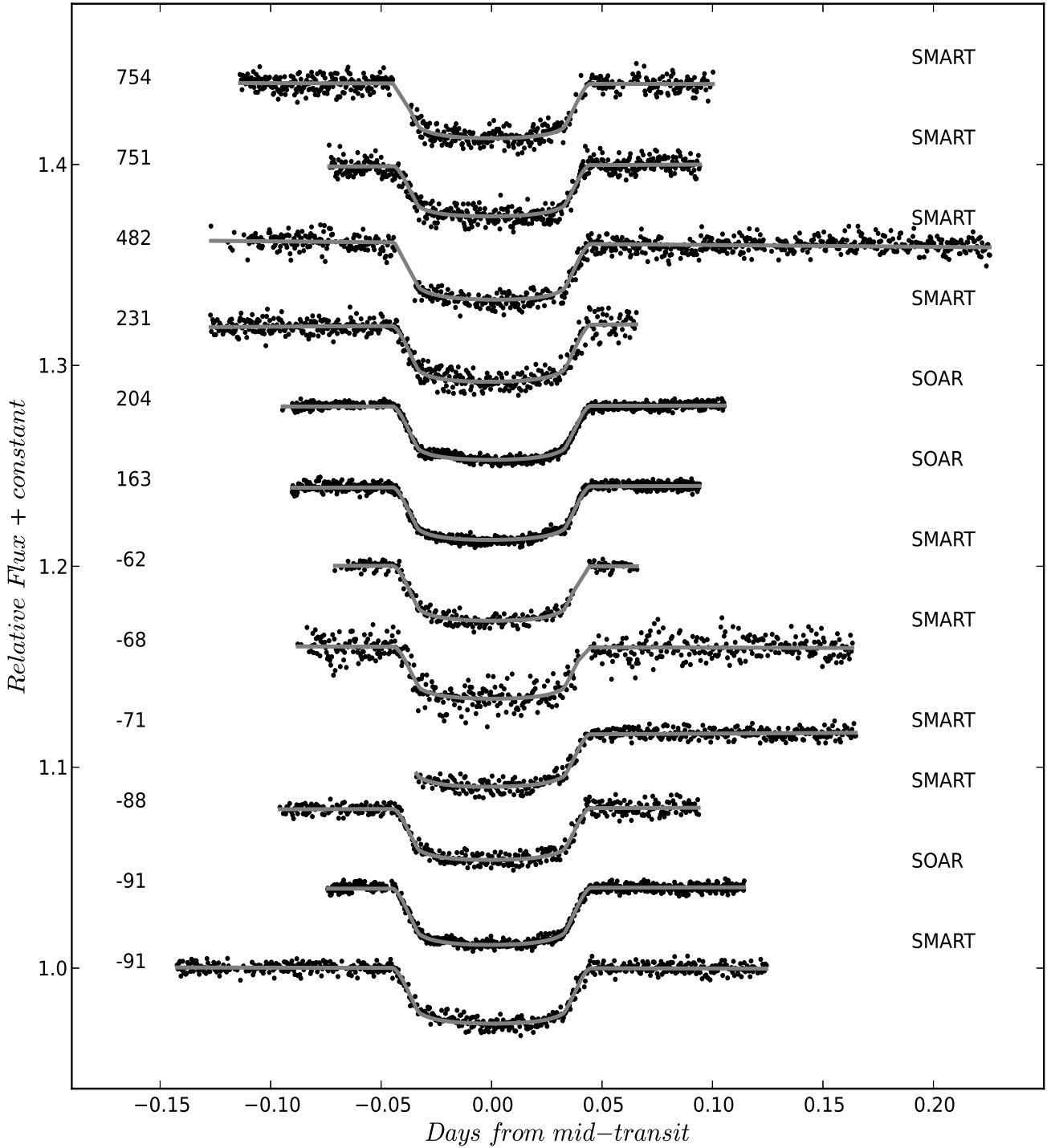}
\caption{Light curves of the twelve new transits of WASP-4b presented
  in this work after the detrending described in Section
  \ref{reduccion}. The solid lines show our best model fits using
  TAP. The epoch number is indicated to the left of each light curve
  and the telescope on the right. }
\label{lc.miscurvas}
\end{figure*}

%% FIGURE 3
\begin{figure*}
\vspace{22.cm}
%\special{psfile=wasp4.lc.publicadas.eps hscale=45 vscale=58 angle=0 hoffset=-10 voffset=-110}
\includegraphics{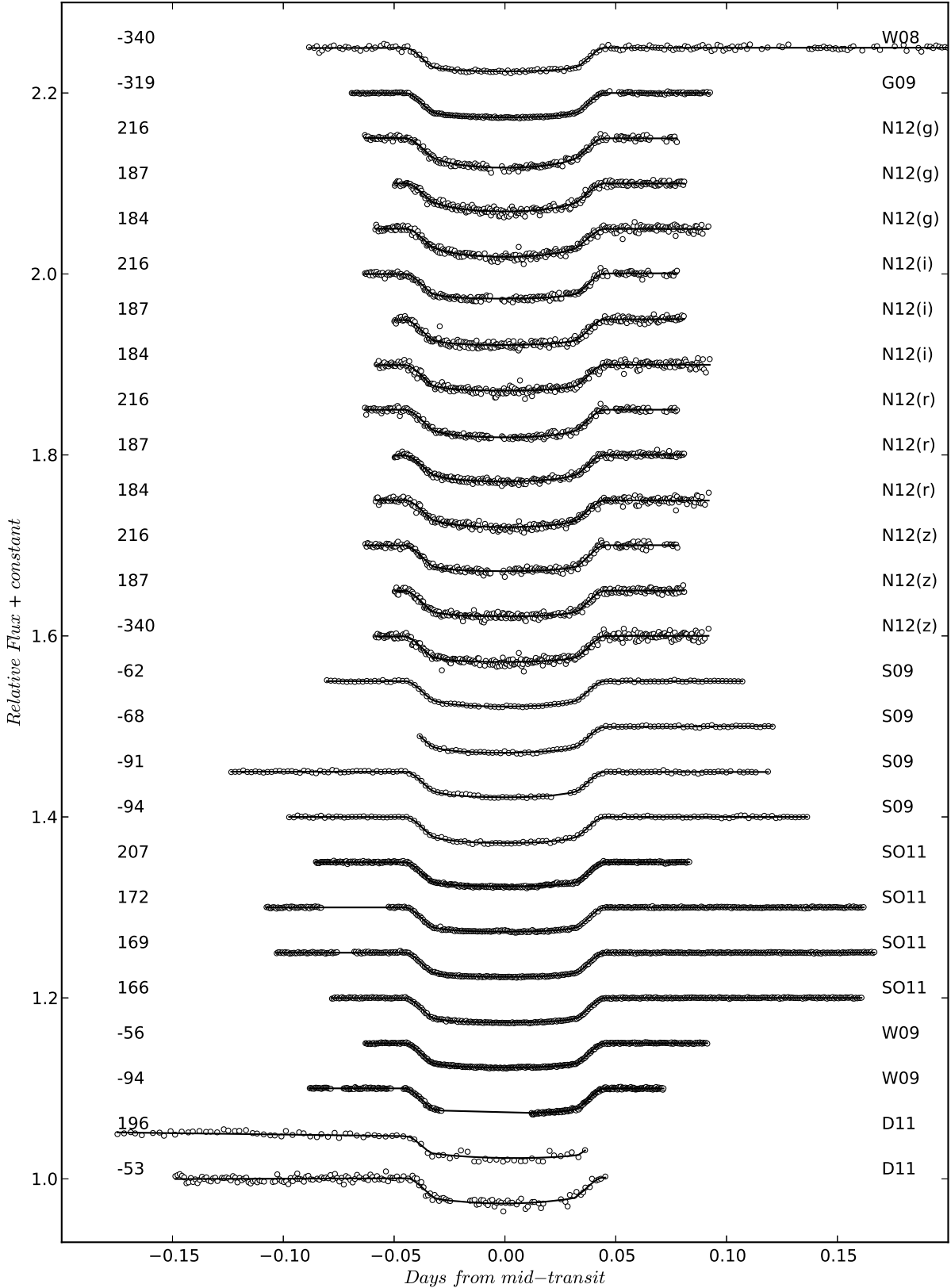}
\caption{Light curves of the twenty-six transits of WASP-4b available in
  the literature.  The solid lines show our best model fits using
  the TAP. The epoch is indicated on the left of each light
  curve. The author of each light curve is indicated on the righ.}
\label{lc.otrascurvas}
\end{figure*}

We fitted the twelve new transit light curves presented in this work
and the other 26 available light curves from W08, G09, W09, S09, D11,
SO11 and N12\footnote{The W08 data are available in the on-line
  version of the article in the ApJL website.  The W09 and S011 data
  are available in the on-line material from the S011 publication on
  ApJ. The S09 and N12 data are available at the CDS
  (http://cdsweb.u-strasbg.fr). The D11 and G09 data were provided by
  the author (private communication).}.  To perform the modelling we
grouped all the light curves observed with the same filter in a given
telescope (we treated all our light curves as observed with the same
filter/telescope).  As we described below, this allows us to fit for
the linear limb-darkening coefficient ($\mu_1$) using all the light
curves simultaneously as was done in \cite{Hoyer-wasp5-2012}.
Therefore, the first group is composed by our ten \textit{I-band}
light curves (from 2008 to 2010) and the second by our two
\textit{Cousins-R} light curves (the 2011's transits).  A third group
is formed by the six SO11 \textit{z-band} light curves (which includes
the two transits observed by W09) and the S09 light curves form the
fourth group (\textit{Cousins-R}). We fit the {\it V}- and {\it
  R}-light curves of D11 separately.  Finally, we fit the twelve light
curves of N12 in four groups (three light curves for each of the four
filters).

We fit each group independently, leaving as free parameters for each
light curve the orbital inclination ($i$), the planet-to-star radius
ratio ($R_p/R_s$) and the central time of the transit ($T_c$).  The
orbital period, the eccentricity, and the longitude of the periastron
were fixed to the values $P= 1.33823326 ~days$ (from D11), $e=0$ and
$\omega = 0$.  We searched for possible linear trend residuals in the
light curves by fitting for the out-of-transit flux ($F_{OOT}$) and
for a flux slope ($F_{slope}$), but did not find any.  The ratio
between the semi-major axis and the star radius, $a/R_{s}$, usually
presents strong correlations with $i$ and $R_p/R_s$.  To break those
correlations and their effects in our results (and
  because we are searching for relative variations in $i$ and
  $R_p/R_s$), we fixed $a/R_s$ in all the light curves to 5.53 (from
  D11). We checked that doing this we are not introducing any
  bias/effect on the rest of the parameters fitted and its errors (for
  details, see \citealt{Hoyer-wasp5-2012}).  In particular we are not
  affecting the mid-times of the transits due to its weak correlation with
  $a/R_s$.  We also fit for a white and red noise parameter,
$\sigma_w$ and $\sigma_r$, respectively, as defined by
\cite{Carter09}.

The coefficients of a quadratic limb-darkening law ($\mu_1$ and
$\mu_2$) in our light curves are strongly correlated, therefore we fixed
$\mu_2$ to 0.32 (based in the tabulated values by \citealt{Claret2000}) and
let $\mu_1$ as free parameter on each group of light curves.

%Therefore, for each group fit we obtained a single value of $\mu_1$
%and a value of $i$, $R_p/R_s$, $\sigma_w$, $\sigma_r$ and $T_c$ for
%each light curve belonging the group.

Therefore, for each light curve fit we obtained a value of $i$,
$R_p/R_s$, $\sigma_w$, $\sigma_r$ and $T_c$ and a single value of
$\mu_1$ for each group.

To fit the transits and derive errorbars for all parameters we ran 10
MCMC chains of $10^5$ links each, discarding the first 10\% of the
results from each chain to minimize any bias introduced by the
parameter initial values.  
%We used a jump rate of $25\%$ for all the fitted parameters.  
The resulting values for each parameter of the
thirty-eight light curves together with their 1$\sigma$ errors are
shown in Table \ref{tabla-parametros-mias} and
\ref{tabla-parametros-otras}.  The data and the best models fits for
all twelve light curves presented in this work are illustrated in
Figure \ref{lc.miscurvas}.  The same is shown in Figure
\ref{lc.otrascurvas} for the twenty-six light curves from the
literature.

\begin{figure}
\vspace{10.cm}
\includegraphics{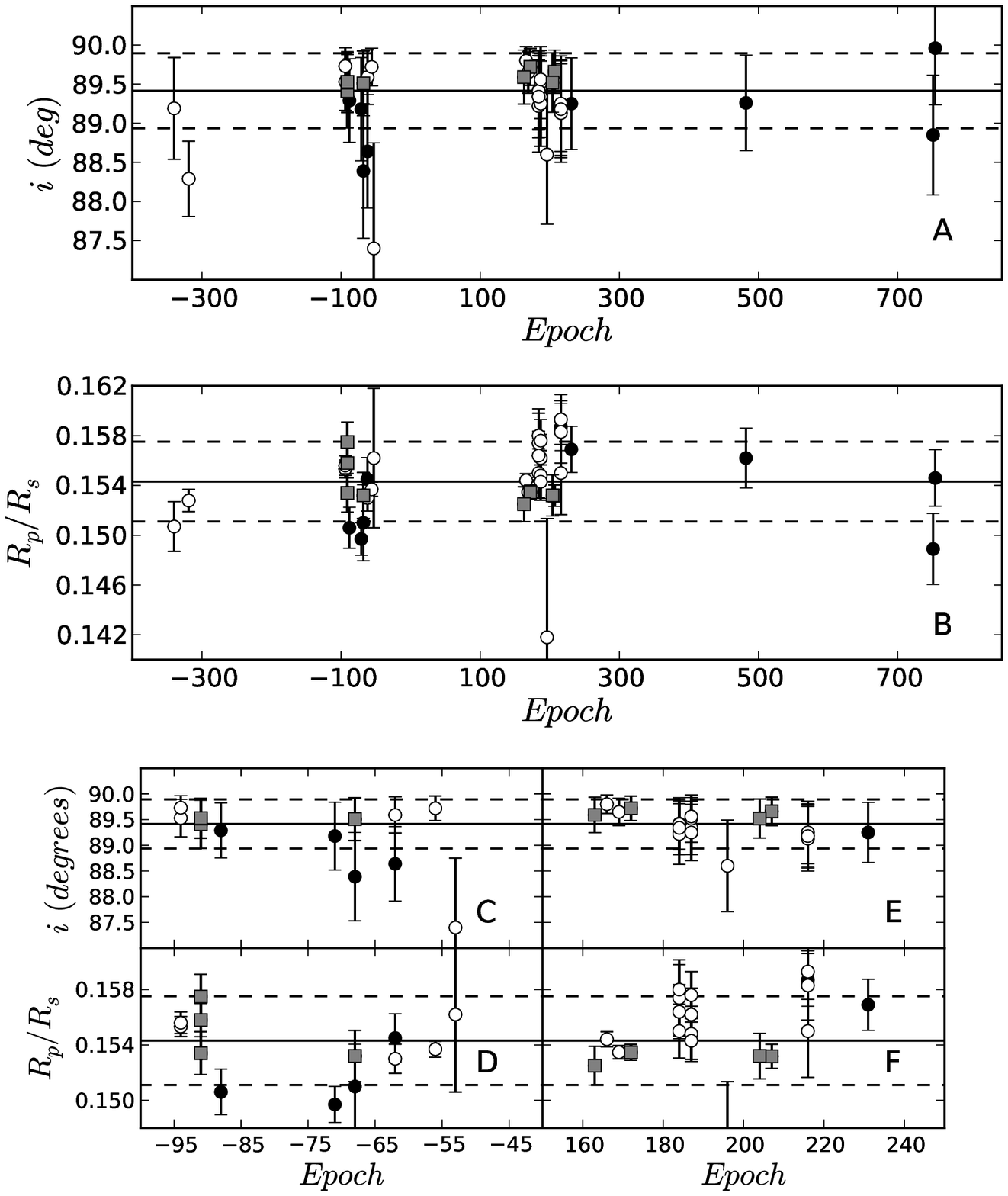}
%\special{psfile=wasp4.figure3.eps hscale=45 vscale=48 angle=0 hoffset=-10 voffset=-40}
\caption{The orbital inclination (\textbf{Panel A}) and the
  planet-to-star radii ratio (\textbf{Panel B}) we derived for the
  twelve transit we present in this work (solid symbols) and for the
  twenty-six transits available in the literature (open symbols), as a
  function of the transit epoch. The parameters derived from transits
  with evidence of star-spot occultations are shown with gray squares.  %In general, for these transits we obtained higher  inclination values.  
  The solid and dashed horizontal lines represent
  the weighted average and its $\pm1\sigma$ errors,
  respectively. There is no evidence of variations on these parameters
  for the time span of the observations. In the bottom panels we show
  a zoom of the diagrams around the -70th (\textbf{Panel C} and
  \textbf{D}) and 200th epochs (\textbf{Panel E} and \textbf{F}).}
\label{parametros}
\end{figure}

Variations in transit parameters, in particular in $i$ and $R_p/R_s$,
can be attributed to the perturbations produced by an additional body
in the system.  In Figure \ref{parametros}, we plot $R_p/R_s$ and $i$
as a function of the transit epoch, based in the results of the 38
transit fits.  We do not find any significant variations in those
parameters.  As reference, in Figure \ref{parametros} the weighted
  average values and the $\pm1\sigma$ errors of $i$ and $R_p/R_s$
  based on all the light curves results are represented by the solid
  and dashed horizontal lines, respectively.  Our analysis of the
  transit timing is described in detail in section \ref{ttv}. 

 Is worthy to noticing that including the detrending
  parameters in the MCMC during the modelling is the most appropriate
  approach to estimate uncertainties of the transit parameters.  We
  have checked that in our data there is no significant differences by
  doing the detrending before the MCMC fitting.  The TAP version we
  use (v2.104) does not allow to detrend against airmass or X-Y
  position of the centroid or neither perform quadratic regression
  fits of the flux against the time, that is why we decided to check
  for correlations against these parameters before modelling with TAP.
  Also, most of the literature light curves we used to search for
  variations in transit parameters have been already detrended and
  therefore the uncertainties we estimate from them can be compared
  with those obtained from our data when we detrended it before the
  MCMC modelling.

%% FIGURE 4

\begin{table*}
\small
\center
%% \tablewidth{0pt}
\caption{Parameters derived for each of the twelves new transits of
  WASP-4b presented in this work. We modeled each light curve
  individually with TAP while $\mu_{1}$ was fitted simultaneously
  from light curves of the same filter. }
\label{tabla-parametros-mias}
\begin{tabular}{lccccccccc}
\hline
\hline
Epoch & Filter &$T_{c}-2450000$  & $i$ & $R_{p}/R_{s}$  & $\mu_{1}$  & $\sigma_{red}^{a}$& $\sigma_{white}^{a}$ & {\it RMS} & Spot \\

 & & ($BJD_{TT}$) & $(\degr)$ &   &  & &  & (residuals) & Detection$^{b}$ \\
 
\hline

%% % MY LIGHT CURVES
% 2000823 - 1mt
-91 & I & $4701.81280^{+0.00022}_{-0.00023}$ & $89.52^{+0.33}_{-0.44}$ & $0.1558^{+0.0012}_{-0.0012}$ & $0.217 ^{+0.019}_{-0.020}$& $0.0066$ & $0.0017$& $0.0023$ & $\surd$\\ 
&&&&&&\\                                           
% 2000823 - SOAR
-91 & '' & $4701.81303^{+0.00018}_{-0.00019}$ & $89.53^{+0.33}_{-0.44}$ & $0.1575^{+0.0016}_{-0.0016}$ & ''& $0.0101$ & $0.0012$ & $0.0015$ & $\surd$\\ 
&&&&&&\\        
% 20080827 - 1mt                                   
-88 & '' & $4705.82715^{+0.00029}_{-0.00030}$ & $89.29^{+0.48}_{-0.59}$ & $0.1506^{+0.0016}_{-0.0017}$ & ''& $0.0066$ & $0.0022$ & $0.0024$ & $\cdots$\\ 
&&&&&&\\         
%  20080919 - 1mt                                 
-71 & '' & $4728.57767^{+0.00043}_{-0.00042}$ & $89.18^{+0.57}_{-0.75}$ & $0.1497^{+0.0013}_{-0.0013}$ &'' & $0.0046$ & $0.0021$ & $0.0023$ & $\cdots$ \\ 
&&&&&&\\                                          
% 20080923 - 1mt 
-68 & '' & $4732.59197^{+0.00050}_{-0.00051}$ & $88.39^{+0.96}_{-0.76}$ & $0.1510^{+0.0031}_{-0.0030}$ & ''& $0.0165$ & $0.0034$  & $0.0049$ & $\cdots$ \\ 
&&&&&&\\                                           
% 20081001 - 1mt
-62 & '' & $4740.62125^{+0.00036}_{-0.00035}$ & $88.64^{+0.82}_{-0.63}$ & $0.1545^{+0.0018}_{-0.0017}$ &'' & $0.0058$ & $0.0019$ & $0.0021$ & $\cdots$ \\ 
&&&&&&\\
%       SOAR                                     
163 & '' & $5041.72377^{+0.00019}_{-0.00018}$ & $89.59^{+0.29}_{-0.40}$ & $0.1525^{+0.0014}_{-0.0014}$ &'' & $0.0095$ & $0.0012$ & $0.0014$ & $\surd$ \\ 
&&&&&&\\    
%      SOAR                                       
204 & '' & $5096.59148^{+0.00023}_{-0.00022}$ & $89.53^{+0.32}_{-0.45}$ & $0.1533^{+0.0016}_{-0.0016}$ & ''& $0.0108$ & $0.0022$ & $0.0013$ & $\surd$ \\ 
&&&&&&\\   
%  20091028 - 1mt                                         
231 & '' & $5132.72310^{+0.00041}_{-0.00041}$ & $89.25^{+0.51}_{-0.66}$ & $0.1569^{+0.0018}_{-0.0019}$ & ''& $0.0082$ & $0.0021$ & $0.0034$ & $\cdots$ \\ 
&&&&&&\\                                           
482 & '' & $5468.61943^{+0.00046}_{-0.00046}$ & $89.26^{+0.52}_{-0.70}$ & $0.1562^{+0.0023}_{-0.0025}$ & ''& $0.0166$ & $0.0034$ & $0.0033$ & $\cdots$ \\ 
&&&&&&\\                                           
751 & R & $5828.60375^{+0.00042}_{-0.00041}$ & $88.85^{+0.75}_{-0.78}$ & $0.1489^{+0.0028}_{-0.0029}$ & $0.212^{+0.066}_{-0.067}$& $0.0167$ & $0.0021$ & $0.0033$ & $\cdots$ \\ 
&&&&&&\\                                           
754 & '' & $5832.61815^{+0.00041}_{-0.00042}$ & $88.96^{+0.69}_{-0.76}$ & $0.1546^{+0.0024}_{-0.0023}$ & ''& $0.0141$ & $0.0023$ & $0.0033$ & $\cdots$ \\  

\hline
\hline
\multicolumn{10}{l}{$^{a}$~$\sigma_{red}$ and $\sigma_{white}$ parameters as defined by \cite{Carter09}.}\\
\multicolumn{10}{l}{$^{b}$ Detections of spot-crossing events during transits ($\surd$: positive detection, $X$: negative detection, $\cdots$: inconclusive). See Section \ref{spots} for details.}\\
\end{tabular}
\end{table*}

% OTRAS -------------------

\begin{table*}
\small
\center
\caption{Parameters derived for each of the 26 transits of WASP-4b reanalized
  in this work.  We modeled each light curve
  individually with TAP while $\mu_{1}$ was fitted simultaneously
  from light curves of the same filter of a given telescope. }
\label{tabla-parametros-otras}
\begin{tabular}{lccccccccc}
\hline
\hline

Epoch & Author$^{a}$, &$T_{c}-2450000$  & $i$ & $R_{p}/R_{s}$  & $\mu_{1}$  & $\sigma_{red}^{b}$& $\sigma_{white}^{b}$ & {\it RMS} & Spot \\
 & Filter & ($BJD_{TT}$) & $(\degr)$ &   &  & & & (residuals) & Detection$^{c}$ \\
%\hline

% Wilson
\hline
-340 & W08, {\it R} &$4368.59279^{+0.00033}_{-0.00032}$ & $89.19^{+0.57}_{-0.73}$ & $0.1507^{+0.0020}_{-0.0020}$ & $0.219^{+0.073}_{-0.078}$ & $0.0036$ & $0.0015$ & $0.0016$ & $\cdots$\\                                  
% Gillon
\hline
-319 & G09, {\it z} &$4396.69576^{+0.00012}_{-0.00012}$ & $88.29^{+0.47}_{-0.49}$ & $0.15279^{+0.00094}_{-0.00085}$ & $0.253^{+0.030}_{-0.037}$ & $0.0027$ & $0.0005$ & $0.0006$ & $X$\\                                          
% Southworth
\hline
-94 & S09, {\it R} & $4697.79788^{+0.00013}_{-0.00013}$ & $89.53^{+0.32}_{-0.41}$ & $0.15533^{+0.00072}_{-0.00072}$ & $0.333^{+0.017}_{-0.017}$& $0.0010$ & $0.0006$ & $0.0008$ & $X$ \\ 
&&&&&&\\                                                                                      
-91 & '' & $4701.81234^{+0.00026}_{-0.00026}$ & $89.41^{+0.41}_{-0.53}$ & $0.1534^{+0.0016}_{-0.0015}$ &'' & $0.0037$ & $0.0007$ & $0.0008$ & $\surd$\\ 
&&&&&&\\                                                                                      
-68 & '' & $4732.59188^{+0.00027}_{-0.00027}$ & $89.51^{+0.34}_{-0.49}$ & $0.1532^{+0.0017}_{-0.0020}$ & ''& $0.0025$ & $0.0005$ & $0.0007$ & $\surd$\\ 
&&&&&&\\                                           
-62  & '' & $4740.62118^{+0.00016}_{-0.00016}$ & $89.59^{+0.29}_{-0.41}$ & $0.1530^{+0.0010}_{-0.0011}$ & ''& $0.0020$ & $0.0005$ & $0.0006$ & $X$\\ 

%Winn+Sanchis-Ojeda
\hline

-94& W09, {\it z} &$4697.79817^{+0.00008}_{-0.00009}$    & $89.73^{+0.19}_{-0.28}$ & $0.15560^{+0.00077}_{-0.00079}$ & $0.2027^{+0.0076}_{-0.0076}$ & $0.0027$ & $0.0005$ & $0.0007$& $\cdots$\\ 
&&&&&&\\                                           
-56& '' & $4748.65111^{+0.00007}_{-0.00007}$ & $89.72^{+0.20}_{-0.28}$ & $0.15369^{+0.00057}_{-0.00058}$ &'' & $0.0024$ & $0.0003$ & $0.0005$ & $X$\\

% Dragomir
\hline

-53& D11, {\it V} &$4752.66576^{+0.00067}_{-0.00069}$ & $87.4^{+1.6}_{-1.1}$ & $0.1562^{+0.0053}_{-0.0059}$ & $0.50^{+0.18}_{-0.16}$& $0.0104$ & $0.0022$ & $0.0029$ & $\cdots$  \\ 
&&&&&&\\                                           
196& D11, {\it R}&$5085.88418^{+0.00084}_{-0.00086}$ & $88.6^{+0.79}_{-0.99}$ & $0.1418^{+0.0092}_{-0.0099}$ & $0.42^{+0.20}_{-0.19}$ & $0.0132$ & $0.0012$ & $0.0026$ & $\cdots$\\

% Sanchis-Ojeda
\hline

166 & SO11, {\it z} &$5045.73853^{+0.00008}_{-0.00008}$ & $89.8^{+0.14}_{-0.22}$ & $0.15441^{+0.00053}_{-0.00055}$ & $0.2027^{+0.0076}_{-0.0076}$ & $0.0023$ & $0.0004$ & $0.0005$ & $X$\\ 
&&&&&&\\                                           
169 & '' & $5049.75325^{+0.00007}_{-0.00007}$ & $89.65^{+0.24}_{-0.29}$ & $0.15347^{+0.00049}_{-0.00047}$ & ''& $0.0018$ & $0.0004$ & $0.0005$ & $X$\\ 
&&&&&&\\                                           
172 &'' &$5053.76774^{+0.00009}_{-0.00009}$ & $89.72^{+0.19}_{-0.28}$ & $0.15346^{+0.00058}_{-0.00058}$ & '' & $0.0026$ & $0.0004$ & $0.0005$ & $\surd$\\ 
&&&&&&\\                                           
207 & ''&$5100.60595^{+0.00012}_{-0.00012}$ & $89.66^{+0.23}_{-0.32}$ & $0.15318^{+0.00086}_{-0.00087}$ & ''  & $0.0043$ & $0.0004$ & $0.0007$ & $\surd$ \\

% Nikolov g
\hline
184 & N12, {\it g'} & $5069.82676^{+0.00031}_{-0.00030}$ & $89.22^{+0.54}_{-0.64}$ & $0.1550^{+0.0019}_{-0.0020}$ & $0.598^{+0.029}_{-0.031}$ & $0.0061$ & $0.0020$ & $0.0026$ & $\cdots$\\                                          
&&&&&&\\  
187 & '' &$5073.84108^{+0.00028}_{-0.00029}$ & $89.34^{+0.47}_{-0.57}$ & $0.1548^{+0.0020}_{-0.0020}$ & '' & $0.0056$ & $0.0017$ & $0.0023 $ & $\cdots$\\         
&&&&&&\\                                           
216& '' & $5112.65009^{+0.00032}_{-0.00033}$ & $89.13^{+0.58}_{-0.68}$ & $0.1593^{+0.0020}_{-0.0020}$ & ''  & $0.0001$ & $0.0015$ & $0.0018 $ & $\cdots$\\     
% Nikolov i
\hline
184& N12, {\it i'} & $5069.82617^{+0.00038}_{-0.00038}$ & $89.42^{+0.41}_{-0.60}$ & $0.1574^{+0.0026}_{-0.0029}$ & $0.238^{+0.037}_{-0.040}$ & $0.0$ & $0.0021$ & $0.0028$ & $\cdots$\\                                          
&&&&&&\\                                           
187&'' & $5073.84128^{+0.00025}_{-0.00026}$ & $89.52^{+0.34}_{-0.58}$ & $0.1562^{+0.0019}_{-0.0019}$ & '' & $0.0$ & $0.0018$ & $0.0025$ & $\cdots$\\                 
&&&&&&\\                                           
216&'' & $5112.65005^{+0.00048}_{-0.00049}$ & $89.25^{+0.52}_{-0.70}$ & $0.1550^{+0.0034}_{-0.0033}$ & ''  & $0.0020$ & $0.0008$ & $0.0019$ & $\cdots$\\               
% Nikolov z
\hline
184 & N12, {\it z'} & $5069.82670^{+0.00028}_{-0.00027}$ & $89.41^{+0.42}_{-0.55}$ & $0.1580^{+0.0018}_{-0.0018}$ & $0.218^{+0.034}_{-0.035}$ & $0.0$ & $0.0023 $ & $0.0029$ & $\cdots$\\      
&&&&&&\\                                           
187 & ''&$5073.84111^{+0.00023}_{-0.00023}$ & $89.25^{+0.51}_{-0.58}$ & $0.1576^{+0.0017}_{-0.0017}$ & '' & $0.0$ & $0.0016$ & $0.0022$ & $\cdots$\\              
&&&&&&\\                                           
216 & '' &$5112.64986^{+0.00036}_{-0.00039}$ & $89.18^{+0.56}_{-0.68}$ & $0.1583^{+0.0025}_{-0.0025}$ & ''  & $0.0$ & $0.0016$ & $0.0020$ & $\cdots$\\                 
% Nikolov r
\hline
184 & N12, {\it r'} & $5069.82661^{+0.00029}_{-0.00029}$ & $89.34^{+0.46}_{-0.59}$ & $0.1564^{+0.0019}_{-0.0020}$ & $0.390^{+0.025}_{-0.027}$ & $0.0003$ & $0.0019$ & $0.0025$ & $\cdots$\\              
&&&&&&\\                                           
187 & ''&$5073.84114^{+0.00018}_{-0.00018}$ & $89.56^{+0.30}_{-0.44}$ & $0.1543^{+0.0013}_{-0.0014}$ & '' & $0.0002$ & $0.0011$ & $0.0017$ & $\cdots$\\
&&&&&&\\                                                                        
216 & '' & $5112.65005^{+0.00031}_{-0.00031}$ & $89.20^{+0.55}_{-0.67}$ & $0.1587^{+0.0019}_{-0.0019}$ & ''  & $0.0$ & $0.0014$ & $0.0018$ & $\cdots$\\
\hline
\hline
\multicolumn{10}{l}{$^{a}$~Transits from  W08 \citep{Wilson08}, G09 \citep{Gillon.WASP4WASP5.2009}, S09 \citep{Southworth.WASP4.2009}, W09 \citep{Winn.WASP4.2009}, D11 \citep{Dragomir11},} \\
\multicolumn{10}{l}{ SO11 \citep{Sanchis-Ojeda2011} and N12 \citep{Nikolov-wasp4-2012}.}\\ 
\multicolumn{10}{l}{$^{b}$~$\sigma_{red}$ and $\sigma_{white}$ parameters as defined by \cite{Carter09}.}\\
\multicolumn{10}{l}{$^{c}$~Detections of spot-crossing events during transits ($\surd$: positive detection, $X$: negative detection, $\cdots$: inconclusive). See Section \ref{spots} for details.} \\
\end{tabular}
%\medskip
%\end{supertabular}
%\end{multicols*}
\end{table*}
%%%%%%%%%%%%%%%%%%%%%%%%%%%%%

\subsection{Global Analysis}\label{global}

\begin{figure}
\vspace{10.6cm}
\includegraphics{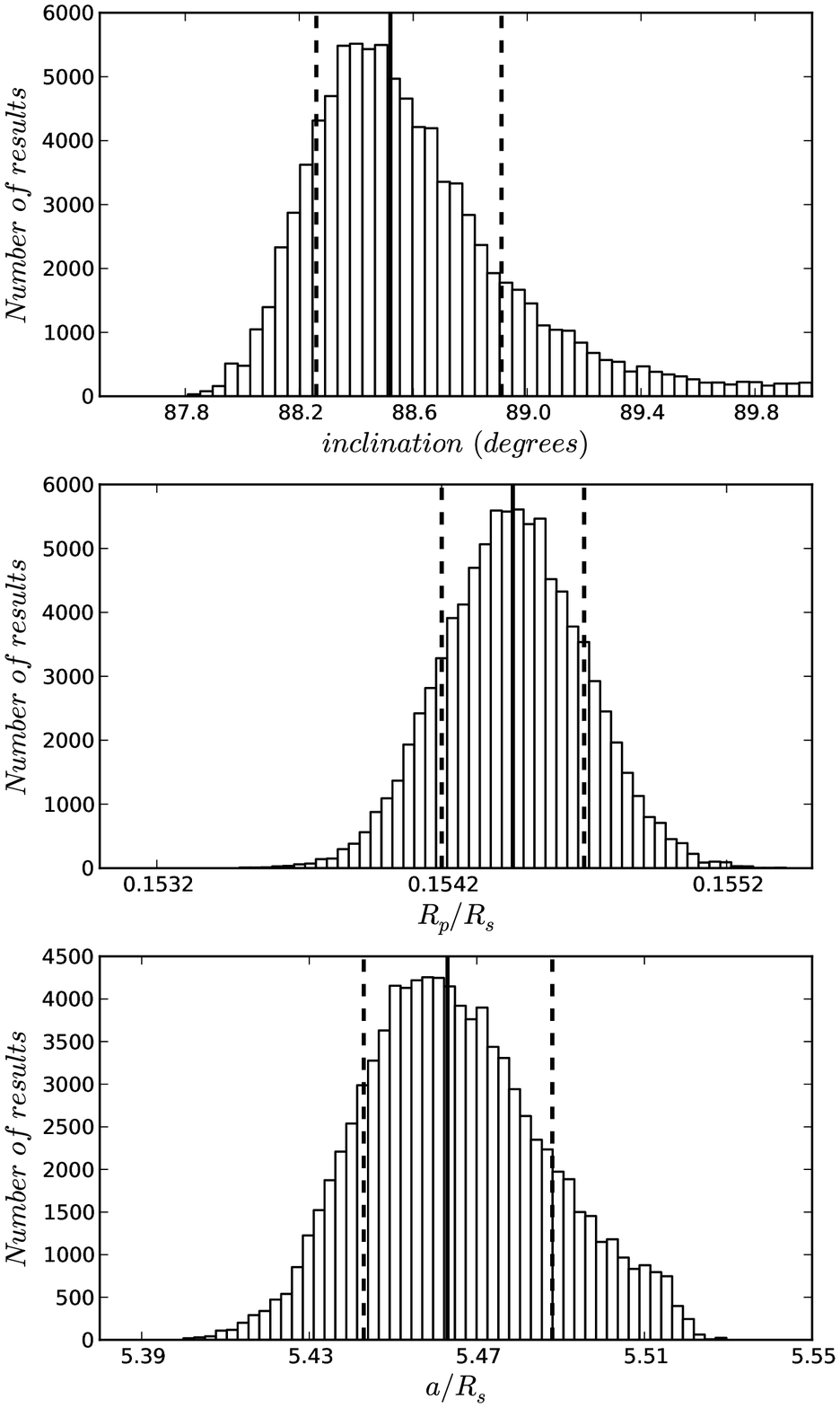}
\caption{ Distributions of the MCMC results of the 10 chains of
  $10^{5}$ links each obtained by fitting simultaneously the orbital
  inclination, $R_{p}/R_{s}$ and $a/R_{s}$ in the 38 available light
  curves of WASP-4.}
\label{histogramas}
\end{figure}

Since we find no evidence of significant variations in the parameters
fitted for each individual light curve, we can model all light curves
simultaneously to improve the determination of $i$, $R_p/R_s$ and
$a/R_s$.  For this, we fitted simulaneausly these parameters in the 38
light curves, while letting to vary on each light curve the transit
mid-times, $F_{slope}$, $F_{OOT}$, $\sigma_{W}$ and $\sigma_{red}$.
We fixed \textit{P} and the linear and quadratic limb-darkening
coefficients for each filter to the values obtained in the previous
section. We used 10 chains of $10^{5}$ links each in the MCMC.  The
resulting values of the MCMC analysis for the simultaneuosly fitted
parameters are shown in Table \ref{tabla-final} and the resulting
distributions in Figure \ref{histogramas}.  Using the resulting value
for $R_p/R_s$ and the value of the star radius ($R_S = 0.907 \pm
0.014~R_{\sun}$) derived by SO11, we obtained an improved radius
measurement for the planet of $R_p= 1.395 \pm 0.022~R_{jup}$, which is
consistent with the most recent radius estimation reported by SO11 and
N12.

%% FIGURE 5

\section{Timing analysis and limits to additional planets} \label{ttv}

%\subsection{WASP-4b}
%% FIGURE 6
\begin{figure*}
\vspace{11cm}
%\special{psfile=wasp4.OC.v3.eps hscale=85 vscale=75 angle=0 hoffset=-12 voffset=-140}
\includegraphics{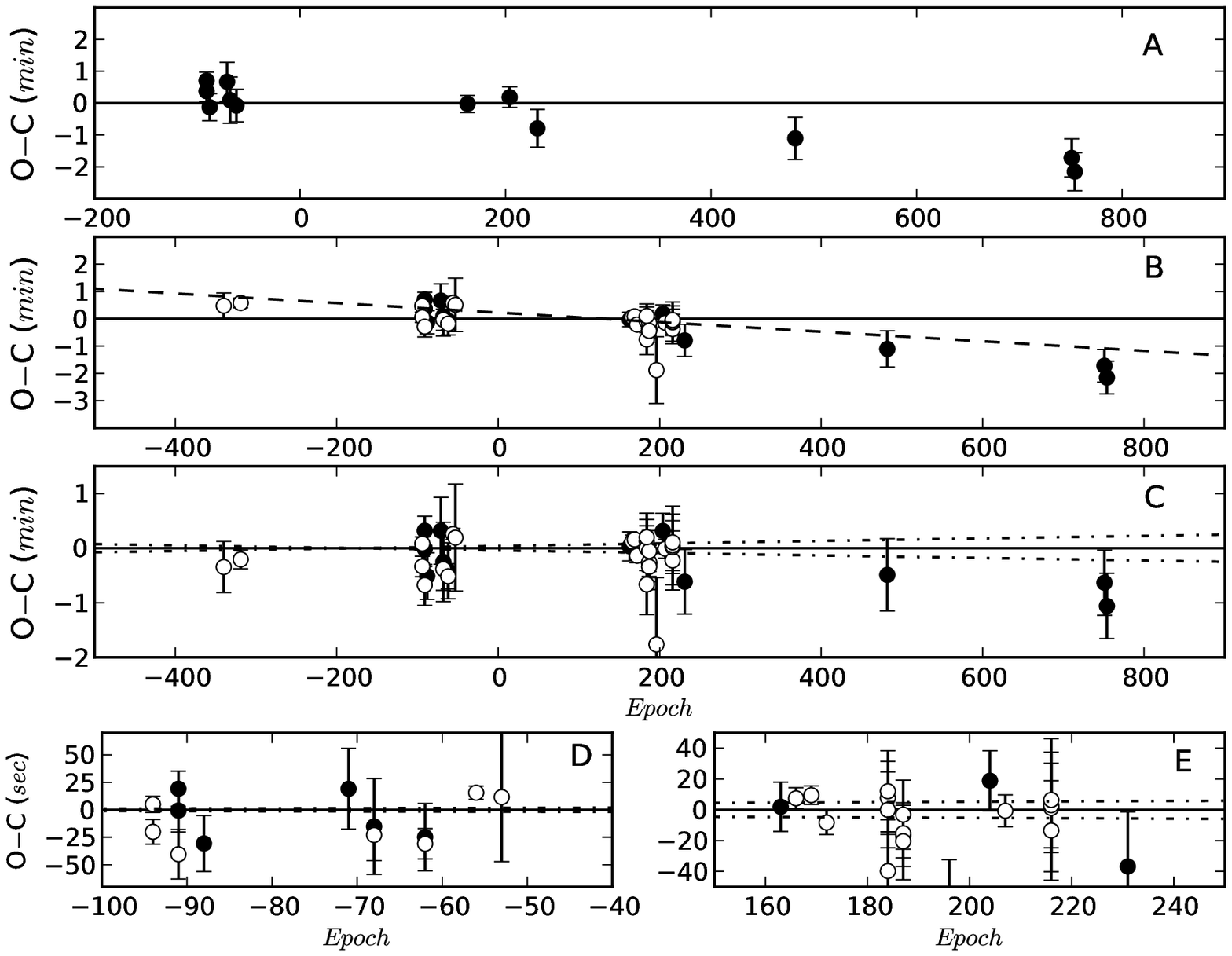}
\caption{$Observed$ $minus$ $Calculated$ diagrams of the WASP-4b's
  transits. \textbf{Panel A:} Timing residuals of the observed new
  twelve transit midtimes presented in this work compared with the
  predicted ephemerides from D11. \textbf{Panel B:} Our twelve $T_{c}$
  (solid circles) combined with the new times derived from W08, G09, W09, S09,
  SO11, D11 and N12 (open circles). A linear trend in the residuals is
  evident. \textbf{Panel C:} If the linear trend is removed, with the
  updated ephemeris equation (shown by the horizontal solid line) the
  $RMS$ of the timing residuals is \rms. The $\pm1\sigma$ levels are
  shown by the point-dashed lines. \textbf{Panel D} and \textbf{E:} we
  show a close view centered in -70 and 200 epoch, respectively. The
  lines are as in Panel C.  An excellent agreement is found in the
  mid-times derived from the common epoch transits and with the
  updated ephemeris equation.}
\label{o-c}
\end{figure*}

The times of our twelve transits and the D11 transits were initially
computed in Coordinated Universal Time (UTC) and then converted to
Barycentric Julian Days, expressed in Terrestrial Time, BJD(TT), using
the \cite{Eastman2010} online
calculator\footnote{http://astroutils.astronomy.ohio-state.edu/time/utc2bjd.html}.
The time stamps of the S09 light curves were initially expressed in
HJD(UT) and have also been converted to BJD(TT).  The same was done
with the transit midtimes obtained from W08 and G09 light curves. No
  conversion was necessary for the light curves reported by SO11
  (which includes the two transits of W09) and the ones reported by
  N12, since they are already expressed in BJD(TT).
%We also used the G09 transit mid-times reported for their \textit{VLT}
%observation and their revisited $T_c$ values for the two transits by W08 (see
%G09, Table 2).  As they suggested we used their reported 'prayer bead'
%errors estimations.  We converted to BJD(TT) those three transit
%mid-times.  
Finally, we did not include the $T_c$ derived from the 2006 and
2007 WASP data since that value is based on the folded transits of
the entire WASP observational seasons, and therefore lacks the
precision required for our timing analysis.

We used the D11 ephemeris equation to calculate the residuals of the
mid-times of the 38 transits of WASP-4b analyzed in this work.  Panel
A in Figure \ref{o-c} shows the $Observed$ $minus$ $Calculated$
($O-C$) diagram of the times for our twelve transits. In panel B we
combined the $O-C$ values of these twelve transits with the new values
derived for the W08, G09, W09, S09, SO11, D11 and N12 (shown as open
circles).  A linear trend is evident in the residuals of all the
transits (represented by a dashed line in panel B).  That trend is
caused by the accumulated error over time in the transit ephemerides.
Once those ephemerides are updated, that trend is removed (panel C),
and the RMS of the transit times residuals is only of $29$ seconds.
The {\it reduced-chi-squared} of the linear regression is
$\chi^2_{red}=1.25$ ($\chi^2=45$ for 36 degrees of freedom). If we
removed the D11's transit with the largest uncertainties (E=196, which
corresponds to an incomplete transit observation) and re-calculated
the linear trend, the RMS of the residuals is only $20$ seconds after
updated the ephemerides ($\chi^2_{red}=1.18$, $\chi^2=41.3$ for 35
degrees of freedom).
%This result is
%  consistent when the central times reported by N12 were included (RMS
%  $= 19$ seconds, $\chi^2_{red}=1.6$, $\chi^2=42.53$ for 26 degrees of
%  freedom).

%Even though a linear regression of the points in the $O-C$ diagram has
%a relative large {\it reduced-chi-squared} of $\chi^2_{red}=1.18$, there
%is enough evidence to conclude that WASP-4b follows a constant orbital
%period.

Once the linear trend is removed (using the linear regression with
$\chi^2_{red}=1.18$) the updated ephemeris equation is:
\begin{equation}
T_{c}=2454823.591924(28)[BJD_{TT}] + 1.33823204(16)  \times E, 
\end{equation}                    
where $T_{c}$~is the central time of a transit in the epoch $E$ since the
reference time $T_{0}$.   The errors of the last digits are shown
in parenthesis.

Panel C in Figure \ref{o-c} shows the resulting $O-C$ values of all
available transits using the updated ephemeris equation. Almost all
the $T_{c}$ coincide with it within the $\pm1~\sigma$ errors
represented by the point-dashed lines.  Despite there is still non
negligible residuals in the timing of $E=751,754$ transits (panel C)
those epochs do not deviate from the updated ephemeris equation by
more than $2.5\sigma$. These points have relative large uncertainties
compare with other epochs and therefore have less weight in the
calculated linear regression.  We found no evidence of a quadratic
function in the $O-C$ values.  We show in panels D and E a close-up of
the $O-C$ diagram around the -70 and 200 epochs, were the transit
observations are more clustered.  All the $T_c$ of the common transits
analyzed in this work are in excellent agreement within the errors.

%We attributed the remaining observed {\it RMS} in the timing residuals
%of the transits to observational/modeling systematics. 

This newly obtained precision permits to place strong constraints in the mass of
an hypothetical companion, particularly in MMR's, as we discuss below.

We used the \textit{Mercury} N-body simulator \citep{Chambers99} to
place upper limits to the mass of a potential perturber in the WASP-4
system, based on our timing analysis of the transits.  A detailed
description of the setup we used for running the dynamical simulations
can be found in \cite{Hoyer-ogletr111-2011} and
\cite{Hoyer-wasp5-2012}.  As a summary, for the simulated perturber
bodies we used circular ($e=0$) and coplanar orbits with WASP-4b.  We
explored a wide range of perturber masses ($0.1M_{\earth}~\leq
M_{pert} \leq 5000 M_{\earth}$) and distances between $0.1 ~AU$ and
$0.06 ~AU$ in steps of $0.001 ~AU$. The semi-major axis steps were
reduced to $0.0005~AU$ near MMRs with the respective transiting body.
The density of the perturber body corresponds to the mean density of
the Earth (for $M_{pert}\leq1M_{\earth}$) or Jupiter (for $\geq 300
M_{\earth}$).  The density was obtained from a linear function that
varies from Earth's to Jupiter's density for all the other $M_{pert}$.

We identified a region of unstable orbits where any orbital companion
experimented close encounters with the transiting body in the time of
the integrations we studied (10 years).  For all the other stable
orbits we recorded the central times of the transits, which were
compared with predicted times assuming an average constant orbital
period for each system.  This period did not deviate by more than
$3\sigma$ from the derived period of each transiting body.  When the
calculated TTV \textit{RMS} approached to 60 seconds we reduced the
mass sampling in order to obtain high precision values ($\leq
1~M_{\earth}$) in the mass of the perturber.

%The results of our dynamical simulations are illustrated in Figure
% for WASP-4b, where we show the mass of the
%hypothetical perturber as a function of the star distance.  The solid
%line in these figures represent upper limits on the mass of a
%perturber that would produce a TTV \textit{RMS} of 20 and 60 seconds
%in the transits of WASP-4b.

\begin{figure}
\vspace{7.cm}
\includegraphics{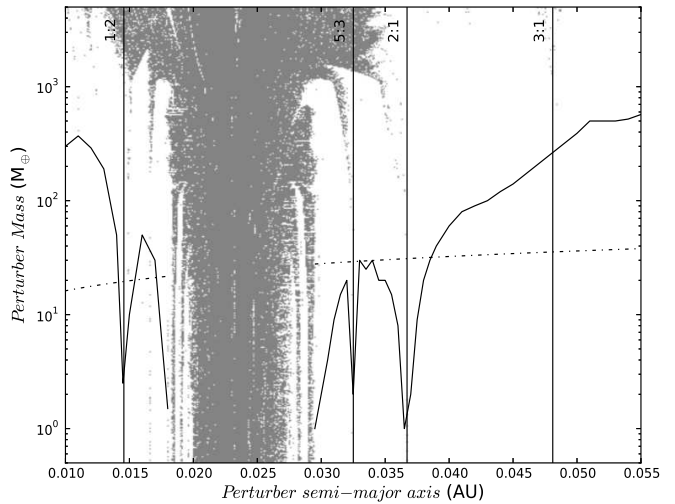}
\caption{Upper mass limits as function of the orbital semi-major axis
  of an hypothetical perturber of the exoplanet WASP-4b, based on
  dynamical simulations done with the Mercury N-body simulator
  (Chambers et al. 1999). The solid line corresponds to perturber
  masses which produce TTVs RMS of $\sim20$ as we measure in our
  timing analysis (see section \ref{ttv}). The point-dashed line shows
  the limits imposed by the radial velocity measurements.  Vertical
  lines mark the locations of MMRs with WASP-4b and the gray region
  indicates MEGNO factor $\langle Y \rangle \ne 2$, i.e.  the region of
  chaotic orbits in the system.}
\label{wasp4.mvsa}
\end{figure}

The results of our model simulations are illustrated in Figure
\ref{wasp4.mvsa}, where we show the perturber mass,~$M_{pert}$
($M_{\earth}$), versus orbital semi-major axis,~$a$ ($AU$), diagram
that places the mass limits to potential perturbers in the system. The
solid line in the diagram indicates the derived upper limits to the
mass of the perturbers that would produce TTV RMS of \rms ~at
different orbital separation. The dashed line shows the perturber mass
upper limits imposed by the most recent RV observations of the WASP-4
system, for which we have adopted a precision of $15~m~s^{-1}$
\citep{Triaud2010}.

In the same figure, we also show the result of calculating the MEGNO
(Mean Exponential Growth of Nearby Orbits) factor $\langle Y \rangle$
\citep{cincottasimo1999,cincottasimo2000,cincottasimo2003} which
measures the degree of chaotic dynamics of the potential
perturber. This technique found widespread application within
dynamical astronomy in studies of stability and orbital evolution, in
particular in extrasolar planetary systems and Solar system bodies
\citep{Gozd2001a,Gozd2001b,Hinse2010}.  Here we use MEGNO to identity
phase space regions of orbital instability in the WASP-4 system. We
calculated $\langle Y \rangle$ on a grid considering 450x400 initial
conditions with the perturber initially placed on a $e=0$ orbit in all
integrations. The transiting planet is located at $a=0.02312~AU$ from
the host star. Each test orbit was integrated for $10^6$ days. For
quasi-periodic orbits $\langle Y \rangle \rightarrow 2.0$ for $t
\rightarrow \infty$ and for chaotic orbits $\langle Y \rangle \ne 2$
(represented by the gray region in the figure).  In theses cases
$\langle Y \rangle$ usually diverges quickly away from 2.0 in the
beginning of the orbit integration. The region close to the transiting
planets is highly chaotic due to strong gravitational interactions
resulting in collisions and/or escape scenarios and coincides with the
unstable region we have identified with \textit{Mercury} code. We
identify the locations of MMRs by the chaotic time evolution at
certain distances from the host star. These coincide with the TTV
sensitivity for smaller masses of the perturber.

For WASP-4b, we found that the upper limits in the mass of an unseen
orbital companion are 2.5, 2.0 and 1.0 $M_{\earth}$ in the 1:2, 5:3
and 2:1 MMRs respectively (vertical lines in Figure \ref{wasp4.mvsa}).
These limits are more strict than the radial velocities constraints,
specially in the MMRs, although by using our approach we can not
reproduce the mass limits derived by N12.

%% Figure 7

\section{Starspot Occultations}\label{spots}

\begin{table}
\small
\center
\caption{New derived parameters for the WASP-4 system
    using the results of the simultaneous modelling of all available
    light curves (Section \ref{global}). The orbital period, P, and
    the reference epoch, $T_{0}$, were determined in the timing
    analysis of the transits (Section \ref{ttv}).}
\label{tabla-final}
\begin{tabular}{lll}

\hline
\hline
Parameter  & Derived Value   & error \\ %& $\chi^{2}_{red}$\\
\hline

%WASP-4b & &  \\  
%\hline

P ($days$)        & $1.33823204$   &   $\pm0.00000016$    \\ %$***$   \\
$T_{0}$ ($BJD_{TT}$)    & $2454823.591924$   &  $\pm0.000028$  \\% & $***$    \\
$i$ (deg)        & $88.52$          &  $+0.39,-0.26 $   \\%&  $***$ \\
$R_{p}/R_{s}$     & $ 0.15445$     &  $\pm0.00025 $  \\% & $***$ \\
$a/R_{s}$        &  $5.463$    & $+0.025,-0.020$ \\% &  \\
$R_{p}$ ($R_{jup}$) &  $1.395$    & $\pm0.022$ \\% &  \\
$e$             & $0$ $^{a}$ & $\cdots$ \\ %&\\ 
%$R_{s}$          & $0.907$ $^{b}$ & $\cdots$ \\ %&\\ 

\hline \hline 
%\medskip
\multicolumn{3}{l}{$^{a}$~This value was fixed in the light curve  modelling.}\\
%\multicolumn{3}{l}{$^{b}$~From \cite{Sanchis-Ojeda2011} and used to derive $R_{p}$.}\\
\end{tabular}
\end{table}

%% FIGURE 8
\begin{figure*}
\vspace{19.cm}
%\vspace{21.4cm}
%\special{psfile=wasp4.spots.gaussian-fits-werrors.eps hscale=70 vscale=78 hoffset=30 voffset=-40}
\includegraphics{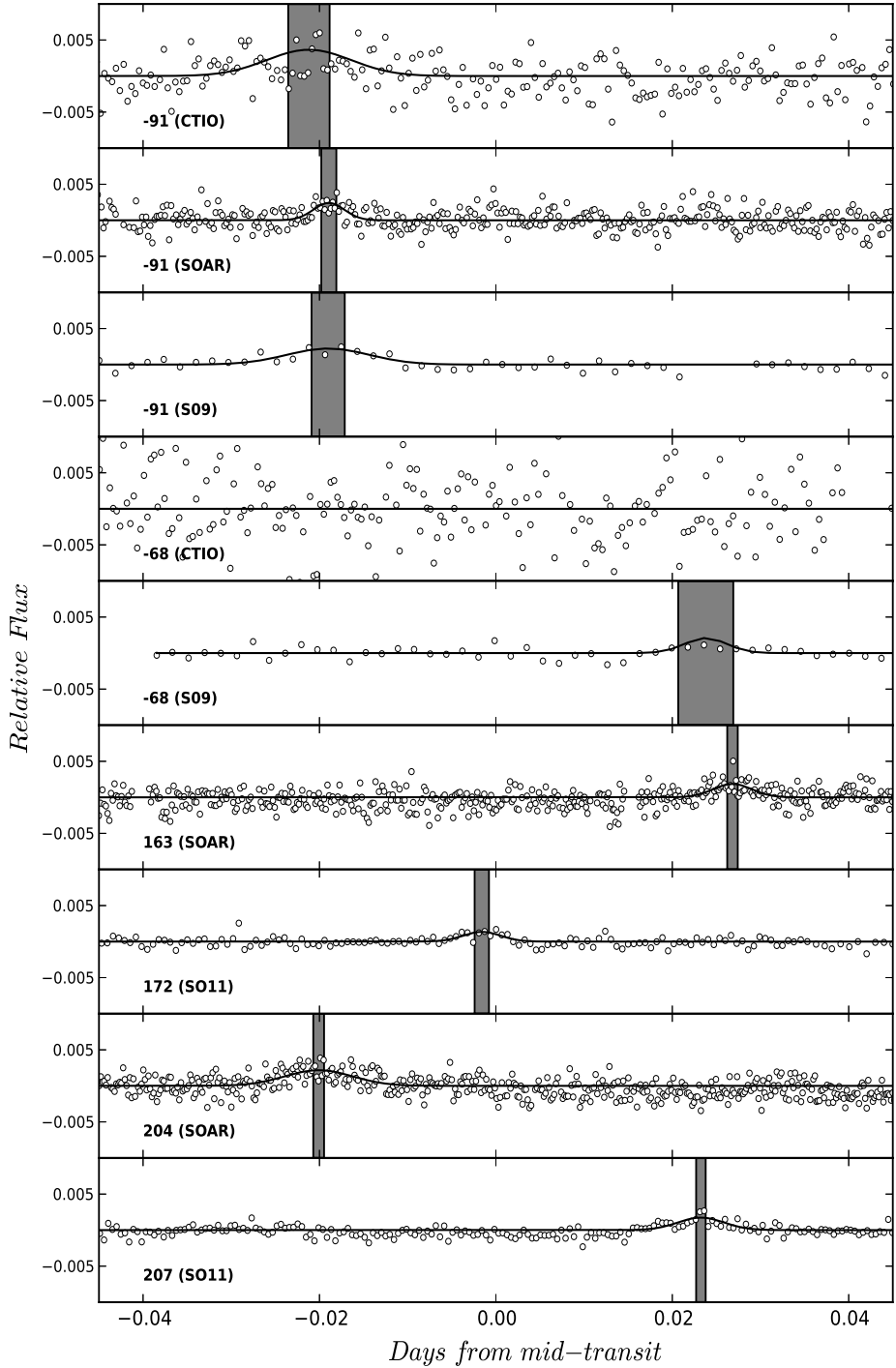}
\caption{Light curves with evidences of star-spot occultations by
  WASP-4b are shown.  The epoch and telescope (or the author) of the
  light curve are shown in the bottom left. The solid lines represent
  the fitted gaussian model and the gray regions are proportional to
  the uncertainties in the timing of the occultations. We confirm the
  detection of the occultation with our -91 epoch's data but in our
  -68 transit we found no evidence of occultation due the low SNR of
  the data. }
\label{wasp4.spots.all}
\end{figure*}

\begin{table*}
\small
\center
\caption{Results of the fitting of the star-spots occultations.  For
  each light curve, the amplitude ($A_{occ}$), the central time of the
  occultation ($T_{occ}$) relative to midtime of the transit ($T_{c}$)
  and the width of the fitted Gaussian function ($\sigma$) are shown.
  The errors were obtained from Monte-Carlo}
\label{tabla-spots}
\begin{tabular}{lcccc}

\hline
\hline
   Epoch   &{\it RMS} & $A_{occ}$ ~(ppm)   & $T_{c}-T_{occ}$ (days)  & $\sigma$ (days) \\
\hline
-91 (CTIO) &  0.0023 & $ 0.0037 \pm 0.0037 $ & $ -0.0211 \pm 0.0023 $ & $0.0048 \pm 0.0027  $\\
-91 (SOAR) &  0.0015 & $ 0.0024 \pm 0.0006 $ & $ -0.0189 \pm 0.0008 $ & $0.0018 \pm 0.0009  $\\
-91 (S09)  &  0.0008 & $ 0.0022 \pm 0.0006 $ & $ -0.0190 \pm 0.0019 $ & $0.0047 \pm 0.0018  $\\
-68 (S09)  &  0.0007 & $ 0.0021 \pm 0.0020 $ & $ 0.024 \pm 0.003    $ & $0.0024  \pm 0.0019 $\\
163 (SOAR) &  0.0014 & $ 0.0019 \pm 0.0003 $ & $ 0.0268 \pm 0.0006  $ & $0.0022 \pm 0.0008  $\\
172 (SO11) &  0.0005 & $ 0.0014 \pm 0.0004 $ & $ -0.0016 \pm 0.0008 $ & $0.0021 \pm 0.0006  $\\
204 (SOAR) &  0.0014 & $ 0.0022 \pm 0.0003 $ & $ -0.0201 \pm 0.0006 $ & $0.0039  \pm 0.0006 $ \\
207 (S011) &  0.0007 & $ 0.0017 \pm 0.0007 $ & $ 0.0232 \pm 0.0005  $ & $0.0026  \pm 0.0009 $ \\
\hline 
%Non-Detections & &&&\\
%-94 (S09) & 0.0008 &&&\\
%-62 (S09) & 0.0007 &&&\\
%-56 (W09) & 0.0005  &&&\\
%169 (SO11)& 0.0004  &&&\\
%166 (SO11)& 0.0005  &&&\\
\hline 
\end{tabular}
\end{table*}

As SO11 previously noted, in some of the light curves during the
transit there are bump-like features or anomalies that can be interpreted
as star-spot occultations by the exoplanet while it crosses in front
the star's disk.  SO11 detected evidence of these occultations in two of the 
light curves they presented and in the two light curves of S09.  Depending on
the \textit{SNR} and the sampling of the data, we also
identified these {\it bumps} in four of our light curves (during three different
epochs).

Using the residuals of the detrended light curves after modelling the
transits (see section \ref{fiteo}), we fit a gaussian function around
each bump feature, initially identified by eye and which we assume are
caused by spots.  In particular we fit for the amplitude ($A_{occ}$),
central time of the spot occultation ($T_{occ}$) and the width
($\sigma$) of the Gaussian which can be used to described the duration
of the occultation event.  To estimate the uncertainties of these
values, before fitting we added to the observational points an amount
of random noise proportional to the \textit{RMS} of the residuals
(without bumps).  We repeated this procedure 10\,000 times for each
light curve and calculated the errors from the width of the resulting
distributions.  In Figure \ref{wasp4.spots.all} we show the results of
those fits for the light curves that present signs of spots in the
stars.  No spots were apparent in the remaining light curves. Our data
of the -91 transit epoch confirms the detection of a spot occultation
in the S09 light curve by SO11.  Given the better sampling and
signal-to-noise ratio (SNR) of our SOAR light curve, we can improve
the central timing of the spot by a factor of three, as illustrated by
the gray vertical bands in Figure \ref{wasp4.spots.all}, which
indicate the error obtained in the central timings of each spots.
However, our CTIO light curve of the transit epoch -68 shows no sign
of spots and we cannot confirm the spot in that epoch in the S09 light
curve.  This can be attributed to the worse SNR of our light curve,
but notice also that the amplitude of the spot occultation that we
obtain in the S09 data is $A_{occ} = 0.0021 \pm 0.0020$.  Therefore
that detection could render spurious.  Among our new data we also
detect two new spots during transit epochs 163 and 204.

In Table \ref{tabla-spots} we show the results of the gaussian fits.
Taking into account the values obtained for $A_{occ}$ and the \textit{RMS}
values of the residuals, no starspot occultations were detected in the
light curves of the epochs: -88, -71, 166 and 169.  These non-detections
(especially the transits closer to those with positive detections) can
be used to constrain the spot's location and/or lifetime.  For
example, it can be argued that during those transits the spot is
located in the non-visible hemisphere of the star, or that the spot
have migrated to different latitudes that those occulted by the
exoplanet's path.

%We noticed that the model fits in these light curves is
%affected by the presence of this protuberance, in particular in the
%estimation of the transit's depth and/or the inclination.  In Figure
%\ref{parametros} we show with solid diamonds the light curves where we
%found these bumps.  While in $R_p/R_s$ we found no any systematic
%trend, we noticed the inclinations values we derived for these light
%curves are, in general, higher than the average.  To properly account
%for these bumps in the light curves fits we need to modeled out the
%star-spot occultation, which is out of the scope of this work.

%% \begin{figure}
%% \vspace{8.cm}
%% \special{psfile=wasp4.lc.spots.eps hscale=45 vscale=50 angle=0 hoffset=-10 voffset=-90}
%% \caption{ Four of our light curves present 'bumps' in the photometry
%%   during the transit, which can be interpreted as evidence of
%%   star-spot occultations.  The location of the 'bump' is marked with a
%%   small vertical line. The epoch and the telescope of the light curve
%%   is indicated in the left-bottom corner of each panel. The fitted
%%   model is shown by the solid line. In some cases, there is an
%%   understimation on the derived transit's depth. \textcolor{red}{agregar -71 ??}}
%% \label{spots}
%% \end{figure}

SO11 found two possible values for the
rotation period of the star, $22.2 ~days$ (based on the spots of their
new light curves), and $25.5 ~days$, based on the S09 light curves.  Both
results are consistent with the constraint of the radial
velocities measurements \citep{Triaud2010}.

We used a simple linear model to estimate the rotational period of
WASP-4 relying only in the timing of the occultations.  We assumed an
aligned system (an assumption consistent with the RM effect
observations and with the SO11's previous analysis of the spots) since
this geometry increases the probability of detecting 
occultations of the same spots by the exoplanet.

We assigned angular coordinates for the location of the spots, such
that at ingress (first contact) and at egress (fourth contact) of the
transit a spot will have a relative angle of $0\degr$ and $180 \degr$,
respectively.  Relative angles in the non-visible hemisphere will
extent between $180\degr$ and $360\degr$. For the
  epochs where no spot occultations were detected we assigned an
  aleatory angle for the location of the spot between 180 and 360
  degrees and assumed larger errors in its timing ($\sim 0.05 ~days$)
  to compensate for the non-detection during transit. Later, as
  explained below, we checked if the rotational period we obtained is
  consistent with the corresponding non-detections. 

Then, we fit the following linear function for the displacement of
the spots:
 
%\[
\begin{equation}
\label{eq_prot}
\delta\Theta = \Theta_0 + \Omega \cdot (T-T_0) , 
\end{equation}
%\] 
where $\Theta_0$ is the relative angle of the spot in an arbitrary
reference time $T_0$, and $\Omega$ is the \textit{displacement rate}
of the spots due to stellar rotation (assuming no migration), in
degrees per days.  To test our model we fit the same spot pairs
analyzed by SO11 (E=172,207), using our newly computed $T_{occ}$ and
obtain a stellar rotation period $P_{rot}=22.7 \pm 0.2~days$,
consistent with the $22.2~days$ period derived by SO11 but
inconsistent with the non-detections of E=166,169.  Our new
occultation central times for the S09 spots (E=-91,-68) give a
$P_{rot}$ of $27.5 \pm 0.5~days$, slightly longer than
  the $25.5 ~days$ period obtained by SO11 but consistent with the
  no-detections in E=-62,-56.  We conclude that, to first order
approximation, our linear model provides a good estimate of the star's
rotational period, given the available data.

  Next we tried to improve the $P_{rot}$ of the star by adding, first, the
  new 204 transit epoch to the 172 and 207 epochs (we assumed the spot
  was the same in the 47 days covered by those epochs). The first
  minimum of the fit to those three epochs gives a $P_{rot}$ of $44 \pm 1
  ~days$ ($\chi^2=10^2$).  That period is almost twice the value derived
  by SO11 and their value of the RV constraints estimated to be $\sim
  (21.5 \pm 4.3~days)\times sin(i_{s})$, where $i_s$ is the inclination
  of the stellar rotation axis with respect to the line of sight.  The
  second minimum of the fit gives a $P_{rot}$ of $22.4 \pm 0.2~days$ but with a
  $\chi^{2}\sim10^3$.  This high values of the $\chi^{2}$ mean that none
  of the obtained values of $P_{rot}$ are consistent with the the
  locations of the spot in the three epochs.  Both solutions deviate
  by more than $10\degr$ in the location of the spot in E=204 event.  The
  second minimun is also inconsistent with the location of the spot in
  E=207 while the first minimum is consistent with it within the
  errors. Furthemore, these solutions are inconsistent with the
  no-detection of the spot during E=166,169.  Probably this is an
  indication that the events occurring in the 172 and 204-207 epoch do
  not correspond to occultations of same star-spot and therefore this
  is a constraint for the lifetime of the spot.  Using only the spot
  $T_{occ}$ of the 204 and 207 epochs we obtained a $P_{rot}$ of $34
  \pm 2~days$.  However, this period do not match the occultation of
  the 172 epoch event, supporting the idea that the occultations in
  E=172 and E=204,207 are over different spots ($\sim43~days$ have
  elapse between the E=204 and E=172 transits).  Nevertheless, is very
  likely that WASP-4b have occulted the same spot in E=204,207 due to
  time span between this events is only of $\sim 4 ~days$. 

To explain why the $P_{rot}=34~days$ is above the limits of the RV's
measurements we can use the observed decrease in the amplitude and
duration of spots between the 204 and 207 epochs.  One possibility is
that we are evidencing a change in the rotational speed of the spots
over the average rotation of the star.  The spot can be migrating to
higher latitudes in the star surface, decreasing by this way the
projected area that is being occulted by the planet, and if WASP-4
presents differential rotation (like the observed in the Sun), the
relative displacement rate of the spot over the stellar surface could
vary.  Therefore the increment in the rotation period we observed
compared with the values of the RV's constraints and the rotational
period we estimated using -91 and -68 data, can be indicative of the
spot migration to latitudes with lower rotational periods.  To fully
support this argument we would need a more dense sample of transit
observations using the same filters and observing configurations.  A
summary of the results of this section is presented in Table
\ref{tabla-prot}.

With the same analysis, if we fit the 163 and 172 epoch's
occultations, the minimum we found corresponds to a rotational period
of $\sim13\pm0.2~days$ (also consistent with the no-detections of the 166
and 169 epochs) but again this period is far below the range indicated
by the RV's measurements and is a evidence that those occultation
events are over different spots.

Of course, a more detailed model is necessary to constrain strictly
the rotation period of the star.  This model has to take into account
other parameters such the amplitude and duration of the spot
occultations, the relative angle between spin axis of the star axis
and the orbital axis of the planet, the lifetime of the spots, etc.

\begin{table}
\small
\center
\caption{Results of the fitting of the rotational period of WASP-4
  using different spot occultation events (indicated with bold
  numbers) and no-detection of occultations during transit. As
  reference, in the second column we indicate the time span between
  the first and last epoch used in the minimization. In the last
  column is indicated if the fitted $P_{rot}$ is consistent with the
  no-detections (No-D).}
\label{tabla-prot}
\begin{tabular}{lclll}

\hline
\hline
   Epochs   &  time span & $P_{rot}$ & $\chi^{2}$ & No-D?  \\
            &  $(days)$  & $(days)$ &  & \\
\hline
  166, 169, \textbf{172}, \textbf{207} & 55 & $22.7\pm 0.2$  & $10^{-5}$  &  no \\
  \textbf{-91}, \textbf{-68} , -62, -56 & 47  & $ 27.5 \pm 0.5 $  & $10^{-7}$  & yes  \\
  166, 169,\textbf{172}, \textbf{204}, \textbf{207} & 55 & $44 \pm 1$  & $10^{2}$  &  no\\
  166, 169,\textbf{172}, \textbf{204}, \textbf{207} & 55 & $22 \pm 0.4$  & $10^{3}$  & no \\
  \textbf{204}, \textbf{207} & 4 & $34 \pm 1$  & $10^{-8}$  & $\cdots$  \\
  \textbf{163}, 166, 169, \textbf{172} & 12 & $ 13 \pm0.2  $  & $10^{-8}$  & yes   \\
\hline 

\end{tabular}
\end{table}

\section{Summary} \label{conclusiones}

We present twelve new transit epoch observations of the WASP-4b
exoplanet.  These new transits observed by the \textit{TraMoS} project
were combined with all the light curves available in the literature
for this exoplanet. It is worth noticing that the analysis and
modelling we perform in Section \ref{fiteo} was done over detrended
light curves (both for the TraMoS and literature data). This does not
significantly affect the results in this paper because the system
does not show TTVs. However, the light curve analysis should be done,
whenever possible, including detrending coefficients as part of a
global parameter fit. Therefore, we encourage authors to provide the
raw light curves data in the publications to allow future homogeneous
analyses of different datasets of a given planet. With an homogeneous
modelling of all this data (the new presented here and the previously
published) we performed a timing analysis of the transits.  Based in
the \textit{RMS} of the $O-C$ diagram of about \rms ~we confirmed that
WASP-4b orbits its host star with a linear orbital period.  We updated
the ephemeris equation of this planet and also refined the values of
the inclination of the orbit and the planet-to-star radii ratio.
Also, during the transits we detected small anomalies in the relative
flux of four of the transits (of three different epochs) presented in
this work.  As \cite{Sanchis-Ojeda2011} previously noted we identified
these anomalies as stellar spot occultations by the planet.  With a
simple modelling using the timing of these occultations we estimated
the rotational period of the star.  With the timing of the events we
are more confident correspond to occultations of a same spot allowed
to propose the rotational period is about 34 days.  Since this value
deviates from the limits imposed by the radial velocities measurements
a further modelling that include spot migration and star differential
rotation is needed.  A monitoring of more consecutive transits will be
necessary to allow for new spots detections and to do a better
constrain in the rotational period of WASP-4.  High cadence light
curves with relative small dispersion are critical on this matter, as
can be seen in our SOAR light curves.

\section{Acknowledgements}

The authors would like thank D. Dragomir and S. Kane for providing
TERMS light curves, M. Gillon for the VLT data and the anonymous
referee for the useful and accurate comments which helped to improve
this manuscript.  S.H. and P.R. acknowledge support from Basal PFB06,
Fondap \#15010003, and Fondecyt \#11080271 and \#1120299. S.H,
received support from ALMA-CONICYT FUND \#31090030 and from the
Spanish Ministry of Economy and Competitiveness (MINECO) under the
2011 Severo Ochoa Excellence Program MINECO SEV-2011-0187 at the IAC.
TCH acknowledges support from KRCF via the KRCF Young Scientist
Fellowship program and financial support from KASI grant number
2013-9-400-00.  TCH wish to acknowledge the SFI/HEA Irish Centre for
High-End Computing (ICHEC) for the provision of computational
facilities and support. V.N. acknowledge partial support by the
Universit\`a di Padova through the \textit{''progetto di Ateneo''}
\#CPDA103591. We thanks the staff of CTIO and SOAR for the help and
continuous support during the numerous observing nights, and
R. Sanchis-Ojeda for helpful comments and discussions.  Based on
observations made with the SMARTS 1-m telescope at CTIO and the SOAR
Telescope at Cerro Pachon Observatory under programmes ID
CNTAC-08B-046,-09A-089,-09B-050,-10A-089 and -10B-066.

\bibliographystyle{mn2ev2}
\bibliography{refs}

\end{document}